\documentclass[sigconf,screen]{acmart}

\renewcommand\footnotetextcopyrightpermission[1]{} 

\pdfoutput=1
\AtBeginDocument{%
  }
\usepackage{fancyhdr}
\usepackage{multirow}
\usepackage{algorithm}
\usepackage{algorithmicx}
\usepackage{algpseudocode} 
\usepackage{multirow}
\usepackage{amsmath,bm}
\usepackage{xcolor}
\usepackage{soul}
\usepackage{graphicx}
\usepackage{booktabs}
\usepackage{amsfonts}
\usepackage{balance}
\usepackage{enumitem}
\definecolor{Seashell}{RGB}{255, 245, 238} 
\definecolor{Firebrick4}{RGB}{255, 0, 0}
\definecolor{lightgreen}{RGB}{232, 251, 245} 
\definecolor{darkgreen}{RGB}{57, 132, 96}
\definecolor{lightblue}{RGB}{200, 225, 243}
\definecolor{darkblue}{RGB}{81, 112, 184}
\newcommand{\redcode}[1]{
  \begingroup
  \sethlcolor{Seashell}
  \textcolor{Firebrick4}{\hl{#1}}
  \endgroup
}
\newcommand{\greencode}[1]{
  \begingroup
  \sethlcolor{lightgreen}
  \textcolor{darkgreen}{\hl{#1}}
  \endgroup
}
\newcommand{\bluecode}[1]{
  \begingroup
  \sethlcolor{lightblue}
  \textcolor{darkblue}{\hl{#1}}
  \endgroup
}
\newcommand{\tool}{\texttt{NLPLego}}

\begin{document}

\title{Intergenerational Test Generation for Natural Language Processing Applications}

\author{Pin Ji}
\email{pinji@smail.nju.edu.cn}
\affiliation{%
  \institution{Nanjing University}
  \city{Nanjing}
  \country{China}
}

\author{Yang Feng}
\email{fengyang@nju.edu.cn}
\affiliation{%
  \institution{Nanjing University}
  \city{Nanjing}
  \country{China}
}

\author{Weitao Huang}
\email{weitao_huang@smail.nju.edu.cn}
\affiliation{%
  \institution{Nanjing University}
  \city{Nanjing}
  \country{China}
}

\author{Jia Liu}
\email{liujia@nju.edu.cn}
\affiliation{%
  \institution{Nanjing University}
  \city{Nanjing}
  \country{China}
}

\author{Zhihong Zhao}
\email{zhaozhih@nju.edu.cn}
\affiliation{%
\institution{Nanjing University}
  \city{Nanjing}
  \country{China}
}


\begin{abstract}
The development of modern NLP applications often relies on various benchmark datasets containing plenty of manually labeled tests to evaluate performance. 
While constructing datasets often costs many resources, the performance on the held-out data may not properly reflect their capability in real-world application scenarios and thus cause tremendous misunderstanding and monetary loss.
To alleviate this problem, in this paper, we propose an automated test generation method for detecting erroneous behaviors of various NLP applications. 
Our method is designed based on the sentence parsing process of classic linguistics, and thus it is capable of assembling basic grammatical elements and adjuncts into a grammatically correct test with proper oracle information.
We implement this method into \tool, which is designed to fully exploit the potential of seed sentences to automate the test generation. 
\tool~disassembles the seed sentence into the template and adjuncts and then generates new sentences by assembling context-appropriate adjuncts with the template in a specific order. 
Unlike the task-specific methods, the tests generated by \tool~have derivation relations and different degrees of variation, which makes constructing appropriate metamorphic relations easier. 
Thus, \tool~is general, meaning it can meet the testing requirements of various NLP applications. 
To validate \tool, we experiment with three common NLP tasks, identifying failures in four state-of-art models. 
Given seed tests from SQuAD 2.0, SST, and QQP, \tool~successfully detects 1,732, 5301, and 261,879 incorrect behaviors with around 95.7\% precision in three tasks, respectively.
The experimental results show that \tool~can efficiently generate tests for multiple NLP tasks to effectively detect erroneous behaviors and evaluate the capabilities of NLP applications without the limitation of benchmarks and tested NLP tasks. 
\end{abstract}





\maketitle

\section{Introduction}
With the development of Deep Neural Networks (DNNs), Natural Language Processing (NLP) technologies have gained significant advancements.
Various NLP applications have been deployed into our daily lives and assist many tasks, such as machine translation for reading foreign literature, sentiment analysis for product recommendations, and Q\&A systems for problem-solving. 
Although some NLP applications have demonstrated capabilities that surpass or equal those of human performance~\cite{zhang2020machine, hassan2018achieving}, they are essentially one kind of software and naturally may suffer from software defects that may cause serious consequences~\cite{Microsoft-chatbot, deriu2021survey,faceabook-wrong-translation, faceabook-thai}. 
Therefore, the quality and reliability of NLP applications are of paramount importance for end-users and urgently require evaluation.
Currently, the standard evaluation paradigm is to estimate the performance using train-validation-test splits~\cite{rajpurkar2016squad}.
To meet the demands of evaluating various NLP applications, researchers have established many benchmarks covering multiple NLP tasks, such as SQuAD~\cite{rajpurkar2016squad}, GLUE~\cite{wang2019glue}, and NIST MT~\cite{NIST-MT}. 
However, due to the limitation of resources, researchers can only sample from usage scenarios and hire crowdsourcing workers to build up these benchmarks.
Thus, the accuracy on the held-out data can not reflect the capability of the NLP applications in real-world application scenarios~\cite{patel2008investigating,recht2019imagenet}. 
Additionally, it is challenging to identify the specific representation of defects and the targeted fixing methods by utilizing solely the aggregate statistical results representing the performance~\cite{wu2019errudite}. 

Despite the pressing need for test generation methods to promote testing effectiveness and efficiency, the nature of DNNs and application scenarios of NLP applications have made testing a challenging task~\cite{ribeiro2022adaptive, chen2021validation}.
The DNNs applied in NLP applications employ a data-driven programming paradigm where all decision logic is learned from a large number of datasets~\cite{larochelle2009exploring,liu2017deep}. 
Consequently, the various test generation methods applicable to traditional software whose internal logic is manually determined by programmers are inadequate for DNN-driven NLP applications. 
Further, constructing proper oracle information for NLP applications, which represents the expected output for the given input, poses a significant challenge for researchers due to the complexity of the low-structured natural language~\cite{barr2014oracle}. 
Especially for Natural Language Generation (NLG) tasks, such as machine reading comprehension, the complexity of grammatical rules, the diversity of sentence patterns, and the multiplicity of word meanings can result in a vast output space, which makes automatically analyzing and understanding the outputs of NLP applications demanding. 
The current evaluation method for NLG tasks employs some accuracy-based metrics (BLEU~\cite{papineni2002bleu}, ROUGE~\cite{lin2004rouge}, etc.) to compute the dissimilarity between the outputs and reference texts (a subset of the correct output space). 
However, studies have demonstrated no significant correlation between these metrics and human evaluation~\cite{wu2016google}.

To assure the quality of NLP applications, researchers have proposed some benchmark-independent evaluation methods. 
For instance, He et al. propose a structure-invariant testing method for machine translation software, which detects translation errors by checking whether the outputs of the MT software satisfy structure invariance~\cite{he2020structure}. 
Chen et al. propose a metamorphic testing method for machine reading comprehension models, including seven metamorphic relations tailored to the machine reading comprehension task (e.g., invertion with the antonymous adjective)~\cite{chen2021validation}.
However, these task-specific testing methods are limited in their scope, and the test generation methods they employed are not easily generalizable due to their focus on specific NLP tasks.
An alternative approach is to use the behavioral testing tool named CheckList~\cite{checklistacl20}, which is designed for multiple NLP tasks and includes widely-used test generation methods. 
CheckList aims to validate whether NLP models adhere to generally accepted basic properties. 
Nevertheless, it requires users to manually confirm the template to ensure the availability of the tests. 

In this paper, we propose an automated test generation method for NLP applications. 
Inspired by sentence parsing, we consider the basic structure of the seed sentence as the derivation template into which the adaptable adjuncts can be inserted. 
In linguistics, a semantically grammatically correct sentence can be assembled from adjuncts and the basic sentence structure consisting of basic grammatical elements~\cite{radford2009introduction, ernst2001syntax}. 
By controlling the number of assembled adjuncts, the semantics and pattern of the sentence can be changed to different degrees~\cite{brown2020syntax}. 
The above process related to identifying grammatical elements and establishing syntactic relations is usually called \textit{sentence parsing}~\cite{friederici2012language}. 
To enhance the diversity of the tests generated, we adopt approaches such as synonym substitution and the use of the masked language model to obtain adaptable adjuncts.
For each seed sentence, this method outputs a derivation tree, which is generated by the assembling step. 
In this tree, the sentences at the different levels do not exhibit consistent variation relative to the seed sentence, and the sentences in the same tree path possess a derivation relation. 
The key aspect of this general-purpose test generation method is that the intergenerational relation established through grammatical elements assembling can be employed in a flexible manner for conducting metamorphic testing to address the absence of oracle information. 
We employ the derivation trees to generate new tests guided by the predefined metamorphic relation. 

We implement this assembling test generation method into an automated testing tool \tool. 
To validate \tool, we conduct experiments on three common NLP tasks, i.e., machine reading comprehension, sentiment analysis, and semantic similarity measures, and identify failures in four state-of-art models. 
Given seed tests from SQuAD 2.0, SST, and QQP, \tool~successfully detects 1,732 incorrect behaviors with 97.3\% precision, 5301 incorrect behaviors with 97.5\% precision, and 261,879 incorrect behaviors with 92.7\% precision. 
The experimental results indicate that \tool~has excellent test generation capability, with the ability to generate 6.5$\sim$71.6 new tests based on one seed test, and the quality of the generated tests is higher than that of the baseline. 
Therefore, we confirm that \tool~combined with metamorphic testing theory can be utilized for testing multiple NLP applications with high efficiency and effectiveness, and can evaluate the capabilities of NLP applications (language understanding, word discrimination, etc.) without the limitation imposed by benchmarks and specific tested tasks. 
The main contributions of this paper are as follows:
\begin{enumerate}
    \item \textbf{Method.} We propose an automated test generation method for NLP applications. 
    This method disassembles a seed sentence into a template and adjuncts, and creates a derivation tree by iteratively mutating and assembling grammatical elements. 
    The derivation relation formed by the assembling step facilitates metamorphic testing and increases the universality of this method.
    
     \item \textbf{Tool.} We implement the assembling test generation method into an automated test generation tool for NLP applications, namely \tool. 
    \tool~employs the derivation trees to generate new tests based on the predefined metamorphic relation and the input format of the tested NLP application. 
    
    \item \textbf{Study.} We experiment with three common NLP tasks and identify failures in four state-of-art models to evaluate the performance of \tool. 
    The results show that \tool~can efficiently generate tests for multiple NLP tasks to detect the erroneous behaviors of the models effectively.

\end{enumerate}

\section{Background}
This section introduces the background knowledge related to the NLP applications and the motivation for proposing the assembling test generation method. 
\subsection{NLP Applications}
Natural Language Processing (NLP) is the subfield of linguistics, computer science, and artificial intelligence concerned with employing computational techniques to learn, understand and generate natural human language data~\cite{hirschberg2015advances}. 
NLP applications can be considered as a hybrid program of the NLP model and supporting code, with the NLP models being the core part that ensures system performance. 
They can be classified according to the NLP tasks they accomplish, and we list three common NLP tasks: 
\begin{itemize}
    \item Machine Reading Comprehension (MRC): MRC is to read and comprehend a text passage to answer questions~\cite{baradaran2020survey}. The input for MRC includes a passage and a question, and the output is the answer. 

    \item Sentiment Analysis (SA): SA is to classify the polarity of a given text~\cite{medhat2014sentiment}. The input for SA can be a word, phrase, sentence, or document, and the output is the predefined sentiment label.

    \item Semantic Similarity Measures (SSM): SSM is to determine the similarity of semantics between a pair of texts~\cite{elavarasi2014survey}. SSM is sometimes regarded as a dichotomous task, with the output being a predefined label indicating whether or not the semantics are the same.

\end{itemize}
Compared to SA and SSM, which organize given text into predefined categories, MRC, which belongs to the natural language generation (NLG) task, is more complex.
NLG tasks require understanding natural language and producing natural language outputs. 
Due to their large output space, it is complicated to construct and check correct outputs manually.
Therefore, we need to comprehensively consider the characteristics of various NLP tasks when designing a test generation method with strong universality. 

\subsection{The Motivation of \tool}

Currently, the standard paradigm for evaluating NLP applications is to use train-validation-test splits to estimate their performance~\cite{rajpurkar2016squad,wang2019glue,NIST-MT}. 
Due to the limitation of resources and the absence of oracle information, researchers can only sample from usage scenarios and hire crowdsourcing workers to build up these benchmarks. 
There is a high probability that tests covering a small number of situations do not trigger potential problems of NLP applications. 
Thus, the standard paradigm can not sufficiently reflect the real performance of NLP applications when facing unpredictable input in real-world application scenarios. 
Furthermore, checking the consistency between the output and the reference is a very one-dimensional way to assess the capabilities of NLP applications, only able to reflect the overall performance of the tested object~\cite{gardner2020evaluating,ribeiro2020beyond}. 
Sometimes, researchers employ tricks to improve the performance of NLP applications, and checking the consistency can not distinguish whether the correct outputs are related to these tricks. 
For example, MRC applications can easily ``guess'' correct answers through memory patterns or keywords, and the consistency can not reflect whether the model truly infers the answers~\cite{gardner2020evaluating}. 
Therefore, we propose the assembling test generation method for NLP applications to alleviate the limitations of benchmarks and tested capabilities. 
This method integrates the ideas of disassembling and assembling to fully exploit the variation potential of the seed test, aims to provide effective tests with diverse sentence patterns and semantics to simulate real-world inputs, and can evaluate the capabilities of NLP applications in language understanding, word discrimination, and other aspects.

\section{Methodology}
This section introduces the assembling test generation method for NLP applications that can disassemble the seed test to generate new tests for improving the testing efficiency and effectiveness. 
\subsection{The Key Idea}
In linguistics, syntactic sentence structure describes the grammatical elements of a sentence and how they are assembled~\cite{brown2020syntax,radford2009introduction}. 
The basic sentence structure consists of basic grammatical elements, including the subject, the predicate, and the object or the complement. 
The omission of basic grammatical elements will lead to grammatical errors. 
Generating sentences with complex syntax can be regarded as assembling their basic sentence structure with multiple grammatical elements that have a modifying or complementary explanatory role. 
The adjunct is a structurally dispensable component of a sentence and can be a single word, a phrase, or an entire clause~\cite{ernst2001syntax}.
The grammatical element is an adjunct if it can be omitted from the original sentence without causing a grammatically incorrect expression~\cite{dearmond1998complements,moravcsik2006introduction}. 
Thus, assembling contextually appropriate adjuncts with the basic sentence structure can transform simple sentences into more complex ones with richer semantics and different pattern. 
The process of acquiring and assembling basic structures and adjuncts is analogous to the process of disassembling and assembling Lego bricks, in which the commonly-used basic components are the base plate and various shapes of bricks. 
The process of assembling Lego blocks is to insert specific kinds of bricks into the base plate in a pre-designed order to construct the complex structures from the basic structure. 
Similarly, we can generate new sentences by assembling the basic sentence structure with adaptable adjuncts in a specific order while keeping the grammar and semantics correct. 

\begin{figure*}[h]
  \centering
  \includegraphics[width=0.95\linewidth]{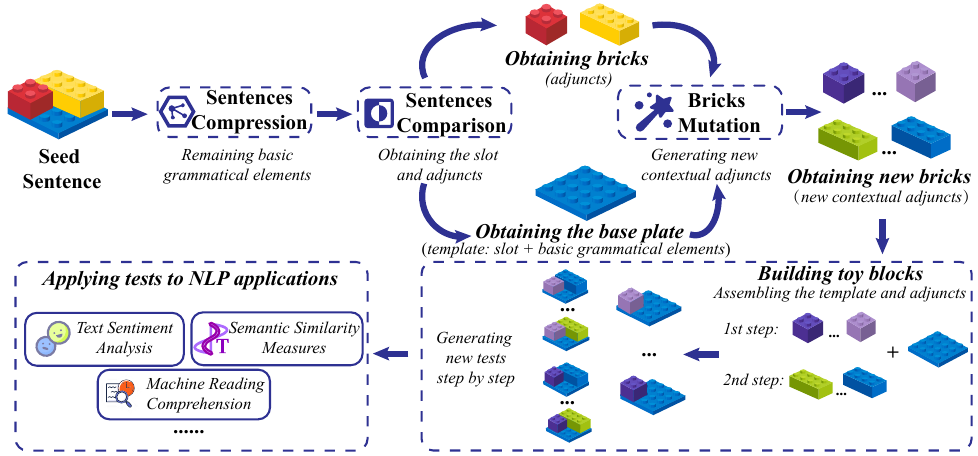}
  \caption{The workflow of~\tool}
  \label{fig:overview}
\end{figure*}

Inspired by the process of sentence parsing described above, we propose the assembling test generation method for NLP applications. 
We employ Lego blocks to present the workflow of this method graphically in Fig.\ref{fig:overview}, and choose a simple seed test containing one sentence to illustrate a running example. 
The assembling test generation method starts by disassembling the seed sentence into the basic structure and adjuncts and then mutating the adjuncts to obtain more adjuncts adapted to the current syntactic structure. 
By inserting adaptable adjuncts in the correct order into the basic structure, we can generate multiple new sentences that possess different patterns and diverse semantics based on the seed sentence. 
Due to the nature of our method, the newly generated sentences have derivation relations and different degrees of variation. 
These characteristics allow us to easily design appropriate metamorphic relations for different NLP tasks to conduct metamorphic testing. 
In the following sections, we introduce the design details of the main steps included in the assembling test generation method. 

\subsection{Sentence Disassembling}
We consider that sentence disassembling is achieved by removing adjuncts while ensuring grammatical correctness. 
The removal process is similar to compressing complex sentences into simpler sentences containing the basic sentence structure. 
Thus, we first identify the adjuncts in the original sentence and remove them to obtain the basic sentence structure. 
During the adjunct identification, we focus on adjectives, adverbs, prepositional phrases, verb phrases, and subordinate clauses in the original sentence, which are often employed as adjuncts in linguistics. 
Subsequently, by comparing the basic sentence structure with the original sentence, we can determine the slots where the adjuncts can be inserted to create the sentence derivation template. 
The derivation template provides guidance for the subsequent assembling steps to ensure the authenticity of the generated sentences. 
We analogize this disassembling process to breaking down the assembled building blocks into multiple bricks and one base plate. 
Formally, given a seed sentence $s$, we formulate the above process of sentence disassembling as follows:
\begin{equation}
    \mathbf{D}(s) \rightarrow \left\{p_{0}, \mathbb{B}\right\}=\left\{s_0 + \textit{pos}(\mathbb{B}), b_{1}, b_{2}, \ldots, b_{n}\right\}
\end{equation}
where $\mathbf{D}$ denotes the sentence disassembling operator, $p_0$ denotes the sentence derivation template containing the basic sentence structure $s_0$ and the slots for inserting adjuncts, $\textit{pos}(\mathbb{B})$ denotes the position of adjuncts in the seed sentence, $\mathbb{B}$ denotes the adjunct set, and $b_i$ denotes the $i_{th}$ adjunct removed by $\mathbf{D}$ from left to right. 

\subsection{Grammatical Elements Assembling}
\label{sec:derivation}
We generate new sentences by assembling the sentence derivation template $p_0$ with an adjunct set $\mathbb{B}$ whose length is $n$. 
We insert the adjunct into the template $p_0$ in order of appearance in the original sentence from left to right. 
For each adjunct inserted, we save the assembling result once.
Thus, multiple sentences with different semantics but structural inclusion relations can be generated based on just one seed sentence. 
According to the definition of adjuncts in linguistics, we can guarantee that each sentence generated by the sequential assembling step is grammatically and semantically correct. 
As shown in the following text box, we can obtain $n$ new sentences and one original sentence after inserting the $i_{th}$ adjunct $b_i$ to the $i_{th}$ slot of the template $p_0$ in order and saving them. 
\vspace{+0.5em}
\begin{center}
\fcolorbox{black}{gray!10}{\parbox{0.95\linewidth}{$s$: On May 3, downtown Jacksonville was ravaged by a fire that started as a kitchen fire.\\
$s_0$: Downtown Jacksonville was ravaged by a fire. \\
$\mathbb{B}=\left\{\text{On May 3, that started as a kitchen fire}\right\}$\\
$p_0$: [1], Downtown Jacksonville was ravaged by a fire [2].\\
$s_1$: Downtown Jacksonville was ravaged by a fire. \\
$s_2$: On May 3, downtown Jacksonville was ravaged by a fire. \\
$s_3$: On May 3, downtown Jacksonville was ravaged by a fire that started as a kitchen fire.
}}
\end{center}
\vspace{+0.5em}

To improve the comprehensiveness of generated tests, we apply the mutation operator $\sigma(b)$ to augment more contextually appropriate adjuncts, which are adapted to the current syntactic structure. 
The new adaptable adjuncts are assembled with the template to cover more situations in application scenarios. 
Each new adaptable adjunct fills the same slot as its seed adjunct. 
Given an initial set $S_0 = \left\{s_0\right\}$, a derivation template $p_0$, and an adjunct set $\mathbb{B}$, we obtain the derivation chain $S_0 \stackrel{P_{0} \bm{+}\sigma(b_1)}{\Rightarrow} S_{1} \stackrel{P_{1} \bm{+}\sigma(b_2)}{\Rightarrow} S_{2} \ldots S_{n-1} \stackrel{P_{n-1} \bm{+}\sigma(b_n)}{\Rightarrow} S_{n}$. 
The derivation template set $P_i$ has the same length as $S_i$. 
By adopting the adjunct mutation operator $\sigma(b)$, the number of generated sentences can be increased exponentially, which is equal to the sum of the lengths of multiple sets $S_0,S_1...,S_n$. 

\begin{figure}[h]
  \centering
  \includegraphics{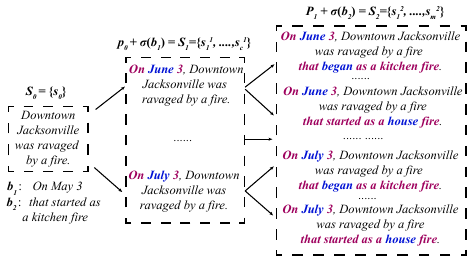}
  \caption{The derivation examples of~\tool}
  \label{fig:example}
\end{figure}

As shown in Fig.\ref{fig:example}, we employ a derivation tree to illustrate the assembling process and the derivation relation between newly generated sentences. 
The root node of the tree is the basic sentence structure of the seed sentence, $s_0$, which serves as the starting point for assembling. 
The leaf nodes are the sentences whose all slots are filled and have the least degree of variation from the original sentence. 
The only difference between the seed sentence and the leaves is in the selection of words, not in the number of words.
Each sentence at the $i_{th}$ level is inserted with one more mutation of adjunct $b_i$ than its parent node at the ${i-1}_{th}$ level. 
Therefore, nodes on each path ${tp}_j$ from the root $s_0$ to the leaf $s^n_j$ are newly generated sentences with the derivation relation.
Formally, given a sentence derivation template $p_0$ and an adjunct set $\mathbb{B}$, we formulate the above assembling process as follows: 
\begin{equation}
\begin{split}
Sent_{Lego} &= \tau\left(p_0, \mathbb{B}\right) = p_0 \bm{+} \sigma(b_1) \bm{+} \ldots \bm{+} \sigma(b_n)\\ &= S_0 \cup S_1 \cup S_2 \ldots S_{n-1}\cup S_{n}
\end{split}
\end{equation}
where $Sent_{Lego}$ denotes the set of new sentences, $\tau$ denotes the assembling operator based on the $p_0$ and the $\mathbb{B}$, and $S_i$ denotes the output of the $i_{th}$ assembling step.
We obtain the set of new sentences as the union of the outputs of each assembling step.

\subsection{Oracle Information Generation}
Given one seed sentence, the sentences in the new sentence set $Sent_{Lego}$ can be classified from two perspectives. 
When classified by the level $i$ of the derivation tree, sentences in each class have the same number of inserted adjuncts, indicating that they have a consistent degree of variation and similar sentence structure. 
When classified by the tree path ${tp}_j$, any two sentences in each class have a derivation relation, indicating that they convey different information and employ different sentence patterns, but possess semantic or structural inclusion relations. 
We combine the above characteristics of newly generated sentences with metaphoric testing theory to address the difficulties in generating the oracle information of tests. 
Metamorphic testing is proposed to address the test oracle problem in test generation. 
It can avoid the possible problems caused by the absence of test oracles through the Metamorphic Relation (MR) definition~\cite{chen2020metamorphic}. 
MR is the necessary property of the target function or algorithm in relation to multiple inputs and their expected outputs~\cite{10.1145/3143561}. 
The violation of MR indicates potential defects in the software. 
Thus, we first construct an appropriate MR according to the characteristics of both the tested NLP task and the sentences in $Sent_{Lego}$. 
Then we employ the new sentence set $Sent_{Lego}$ to generate a new testing set $Test_{Lego}$ in the light of the predefined MR and the input format of the tested NLP task.

\section{IMPLEMENTATION}
We implement the assembling test generation method into an automated testing tool~\tool. 
This section introduces the implementation details, including our design of the disassembling operator $\mathbf{D}$, the assembling operator $\tau$, and the adjunct mutation operator $\sigma$.

\subsection{Template Generation}
The input of sentence disassembling operator $\mathbf{D}$ is a seed sentence, and the outputs are the derivation template $p_0$ and an adjunct set $\mathbb{B}$. 
We employ \textit{spaCy}~\cite{spacy} to obtain the dependency structure of the seed sentence and the shift-reduce parser of \textit{Stanford CoreNLP}~\cite{corenlp} to obtain the constituency structure. 
Dependency reflects the part-part relation between pairs of words in the sentence, and constituency reflects the part-whole relation between a sentence and one of its components~(word, phrase, clause, etc.)~\cite{mazziotta-kahane-2017-extent}. 
Through practice, we find that the dependency is more suitable for extracting adjectives, adverbs, prepositional phrases, and verb phrases, while the constituency is more suitable for extracting subordinate clauses. 
We first extract these components according to the dependency and constituency structure of the seed sentence. 
Then we employ an advanced sentence compression model \textit{SLAHAN}~\cite{kamigaito2020syntactically}, which can generate a simple
sentence that retains basic grammatical elements by omitting redundant components. 
The output of \textit{SLAHAN} is a label sequence consisting of zeros and ones whose length is the same as the number of tokens of the input sentence. 
The token should be retained if the label corresponding to the token is 1; otherwise, it should be removed. 
We calculate the \textit{attention weights} of components based on the outputs of \textit{SLAHAN}, which represent the grammatical importance of components to the seed sentence. 
We regard the number of the one label in the label sequence corresponding to these components as their \textit{attention weights}.
If the \textit{attention weight} of the component is greater than half of the number of tokens it contains, the component is grammatically important and should be retained; otherwise, the component is considered an adjunct and can be removed. 
Once the adjunct set $\mathbb{B}$ is determined, we can obtain the compression result $s_0$ and the $n$ slots, which together form the derivation template $p_0$. 

\subsection{Sentence Generation}
We devote the assembling operator $\tau$ to deriving new sentences based on the derivation template $p_0$ and an adjunct set $\mathbb{B}$. 
When generating the sentences at the $i_{th}$ level of the derivation tree, $\tau$ inserts one mutation of the $i_{th}$ adjunct into the sentence at the ${i-1}_{th}$ level. 
The inserted position is the $i_{th}$ slot in the template $p_0$. 
To improve the comprehensiveness of generated tests, we adopt the idea of fuzzy testing to infer contextually appropriate adjuncts by performing the mutation operator $\sigma$ on the original adjuncts.
In each mutation process, we substitute one word in the adjunct, which can be a noun, verb, adjective, or adverb, and does not belong to the stop-words (commonly-used words, e.g., she, the) provided by \textit{NLTK}~\cite{2004NLTK} to avoid grammatical errors.
We adopt two methods to determine the candidates that can be filled in the position of the substituted word.
The first method is synonyms substitution. 
We search for synonyms of each substituted word in the online lexical database \textit{WordNet} corpus~\cite{Miller1995WordNet}. 
We further refine the search results based on the singular and plural of the substituted noun, the tenses of the substituted verb, and the grade of the substituted adjective. 
This mutation method ensures that the semantics of adjuncts does not change as much as possible. 
The second method is to employ the masked language model (MLM) to predict which words fit in the context of the substituted word~\cite{liu2020survey}. 
First, we employ ``[MASK]'' to replace the words that meet the substitution condition in the $i_{th}$ adjunct to obtain multiple adjuncts with one ``[MASK]''.
Then, we insert these masked adjuncts into the sentences at the ${i-1}_{th}$ level and input these masked sentences into the masked language model. 
The masked language model outputs which words are possible to appear at the masked position. 
The semantics of adjuncts mutated by this method may change but still fit the context, thus enabling generated sentences to cover more semantics. 

\subsection{Metamorphic Relation Construction}
In this paper, we employ \tool~to generate tests for three common NLP tasks: MRC, SA, and SSM. 
Based on the derivation relations and the different degrees of variation in the new sentences, we design different MRs for the aforementioned NLP tasks. 
For MRC, we determine the test oracle through the semantics invariance relation on the outputs of MRC applications. 
Given a paragraph $p$, a question $q$ related to the paragraph $p$, the semantics invariance relation can be extended as follows:
\begin{equation}
    \mathbb{T}\left(p, p^{\prime}\right) \rightarrow \mathbb{T}\left(\mathbf{M}_{mrc}(p, q), \mathbf{M}_{mrc}\left(p^{\prime}, q\right)\right)
\end{equation}
where $p'$ denotes the new paragraph that has the same semantics with $p$, $\mathbf{M}_{mrc}$ denotes the MRC application whose output is the predicted answer of the question $q$, and $\mathbb{T}$ denotes the semantics invariance relation. 
The semantics invariance means that the semantics of the predicted answer should not change when reading the paragraphs with the same semantics to answer the same question. 

For SA, we set a directional expectation to address the absence of test oracles for new tests. 
We expect changes in sentiment to conform to certain expectations.
We posit that the sentiment should not become more \textit{negative/positive} if we add a \textit{positive/negative} text component. 
For example, ``I like that pretty girl'' is more positive than ``I like that girl.''
This directional expectation relation has been employed in SA testing~\cite{ribeiro2020beyond}.
When conducting testing, we regard the changes in the probability of the label as the changes in sentiment. 
Given a sentence $s$ and a text component $c$, the directional expectation relation can be extended as follows:
\begin{equation}
     s + c \rightarrow P_{\mathbf{M}_{sa}(c)}(s + c) \uparrow
\end{equation}
where $\mathbf{M}_{sa}$ denotes the SA application whose output is the sentiment label, $P_{\mathbf{M}_{sa}(c)}(s + c)$ denotes the probability of the label $\mathbf{M}_{sa}(c)$ when the input is $s+c$, and $\uparrow$ denotes as an increase of the probability. 
An erroneous behavior is detected when the increase of the probability of \textit{negative/positive} increases above the change threshold when a \textit{positive/negative} component is inserted. 

For SSM, we focus on evaluating the capability of understanding semantics. 
Sentence disassembling can turn a concrete sentence into a more abstract one. 
Concrete sentences contain more detail than abstract ones. Thus their semantics are relatively different. 
Given two sentences $s$ and $s'$ that have the derivation relation, the semantics variance relation can be extended as follows:
\begin{equation}
    {Sem}_s \neq {Sem}_{s'} \rightarrow \mathbf{M}_{ssm}\left(s, s'\right) = 0
\end{equation}
where ${Sem}_s$ and ${Sem}_{s'}$ denote the semantics of $s$ and $s'$, and $\mathbf{M}_{ssm}$ denotes the SSM application whose output is zero or one. 
Zero indicates that the semantics of $s$ and $s'$ are different. 
The transformation relation between abstraction and concretization coincides with the derivation relation in the outputs of~\tool. 

\section{Experimental Settings}
To comprehensively evaluate the performance of~\tool, we select three common NLP tasks as application scenarios of~\tool, i.e., Machine Reading Comprehension, Sentiment Analysis, and Semantic Similarity Measures. 
For each NLP task, we select the core component of the NLP application, the advanced NLP model, as the tested objects and employ widely-used datasets as the seed sets. 
Specifically, this section aims at answering the following four research questions in each application scenario: 
\begin{itemize}
\item  \textbf{RQ1. Efficiency:} How efficient is \tool~in generating tests and detecting erroneous behaviors?
\item \textbf{RQ2. Authenticity:} How authentic are the tests generated by \tool? 
\item \textbf{RQ3. Effectiveness:} How effective is \tool~in detecting erroneous behaviors?
\item \textbf{RQ4. Errors:} Where do False Positives (FP) and False Negatives (FN) come from?
\end{itemize}

\subsection{Datasets and Models}
We select three widely-used datasets as seed sets, i.e., SQuAD 2.0~\cite{2018Know}, the Stanford Sentiment Treebank (SST)~\cite{socher2013recursive}, and Quora Question Pairs (QQP)~\cite{wang2019glue}. 
Table~\ref{tab:dataset-info}~shows the statistical information of these datasets. 
SQuAD 2.0 contains over 100,000 question-answer pairs on over 500 articles in SQuAD~\cite{rajpurkar2016squad} and over 50,000 unanswerable questions. 
These confusing unanswerable questions were written by crowdsourcing workers. 
SST is a corpus with fully labeled parse trees, and all the sentences it includes are movie reviews. 
Thus, it can be used for analyzing the compositional effects of sentiment in one sentence. 
SST includes 215,154 unique phrases extracted from 11,855 parse trees. 
QQP consists of question pairs collected from the community question-answering website Quora. 
It includes over 400,000 lines of potential question duplicate pairs and corresponding labels and is often applied for training the model to determine whether sentences are semantically equivalent~\cite{WinNT}.

\begin{table}[htbp]
\small
\caption{Numbers of samples for each dataset.}
\label{tab:dataset-info}
\begin{tabular}{ccccc}
\toprule
\textbf{Dataset}&\textbf{Language}&\textbf{\#Train Set}&\textbf{\#Dev Set}&\textbf{\#Test sets}\\
\hline
SQuAD 2.0&English&130k&12k&8862\\
\hline
SST&English&67k&872&1821\\
\hline
QQP&English&364k&40k&390k\\
\bottomrule
\end{tabular}
\end{table}

We select ALBERT~\cite{lan2019albert}, XLNet~\cite{yang2019xlnet}, RoBERTa ~\cite{liu2019roberta}, and ERNIE 2.0~\cite{sun2020ernie} as the tested NLP models. 
The accuracy of RoBERTa-large and ERNIE 2.0-large finetuned on SST is 94.7\% and 95.3\%, respectively. 
The accuracy of RoBERTa-large and ERNIE 2.0-based finetuned on QQP is 92.0\% and 95.8\%, respectively. 
For MRC, we employ pretrained ALBERT-base and XLNet-large finetuned on SQuAD 2.0, achieving 82.1 and 86.3 F1, respectively. 
These models are widely used in current research and can represent state-of-the-art levels in various NLP tasks.


\subsection{The Settings of Test Generation}
\label{ref:settings}
A complete MRC test consists of a paragraph, a question, and the corresponding answer. 
We randomly select four sentences from the paragraph in the seed test and employ \tool~to reconstruct them. 
The critical point of reconstruction is to maintain their semantics consistency as much as possible. 
Therefore, in the reconstruction process, we choose synonym substitution to mutate adjuncts and regard the leaves of each derivation tree with the minimum degree of variation as the reconstruction results. 
Moreover, we do not mutate the adjuncts related to answers and questions. 
The seed set is the development set of SQuAD 2.0. 
To improve the efficiency of \tool, Beam Search is adopted for pruning the derivation trees. 
Beam search is a heuristic search algorithm employed in the NLP field~\cite{tillmann2003word}. 
It chooses the best output for the target variables (e.g., maximum probability). 
We set the target variable as the maximum similarity between word vectors and set the beam size as 4. 

For SA and SSM, we employ the masked language model (e.g., BERT~\cite{kenton2019bert}) to mutate the adjuncts. 
\tool~generates the derivation tree for each sentence in the seed set. 
Unlike MRC, all sentences in the derivation tree are fully utilized during SA and SSM testing. 
We regard all sentences in the derivation tree and the inserted adjuncts as new tests of SST, and set the change threshold to $0.1$.
The test of QQP consists of two questions, so we regard all sentence pairs with the derivation relation in each derivation tree as new tests. 
According to the metamorphic relation we defined, we infer that the semantics of the two sentences having a derivation relation are different. 
For SA, the seed set is the testing set of SST, and for SSM, the seed set is the development set of QQP. 

\subsection{Baseline}
We choose the behavioral testing tool for NLP applications, CheckList~\cite{checklistacl20}, as the baseline because it is a general-purpose testing tool and more similar to \tool~than other task-specific testing tools. 
It contains simple abstractions for writing test expectations and integrates some common perturbations.
Users must manually create new abstractions if they have new testing requirements. 
The user studies show that CheckList can find new and actionable bugs in different NLP tasks. 
We adopt the automated test generation methods in CheckList other than the manual creation. 
Like \tool, we employ the synonym substitutions method integrated in CheckList to generate new tests in MRC testing. 
For SA and SSM, we employ the methods of adding positive/negative phrases, changing the named entity/number/temporal logic, and injecting a negative word integrated in CheckList. 
The number of tests generated by CheckList is related to the structure of the seed sentence and the number of figures and named entities contained in the seed sentence. 
When using CheckList, we select the same substituted range as \tool~and adjust relevant parameters to ensure that the number of generated tests is close to \tool. 

\subsection{Evaluation Metrics}
\label{sec:metric}
\tool~has a strong capability of generating tests, making constructing ground-truth manually for each test a challenging task. 
It is costly in terms of labor and time. 
Thus, we select precision as the evaluation metric, which only requires the number of false positives and true positives. 
The precision of~\tool~can be regarded as the percentage of reported bugs that are truly related to the incorrect behaviors of the tested object. 
We calculate the precision of \tool~employing the following formula: 
\begin{equation}
    \text {Precision}_{\text{bug}}=\frac{\sum_{b \in B}\{\operatorname{report}(b)\}}{|B|}
\end{equation}
where $B$ is the set of suspected bugs reported by~\tool, and $|B|$ is the size of $B$. 
We set $report(b)$ to true if the reported bug $b$ is related to the incorrect behaviors of the tested object and set $report(b)$ to false if it is not. 
To evaluate the performance of checklist, we also employ this evaluation metric.

\subsection{Ground-Truth Construction}
Considering that different people may hold different opinions on the erroneous behaviors of NLP applications, we let multiple judges diagnose the suspected bugs reported by \tool.
With all the seed sets being in English, we hire eight graduate students majoring in English who have been researching English for more than four years. 
Before they start diagnosing, we provide them with a task description and train them to make sure they have fully understood the task. 
For MRC, each suspected bug consists of a paragraph, a question, and the answer predicted by the tested object. 
Judges are required to read and comprehend the paragraph to check if the answer is incorrect. 
For SA, each suspected bug consists of the original sentence, the inserted adjunct, the new sentence, their sentiment label output by the tested object, and the change in the probability corresponding to the sentiment label. 
Judges are required to determine if the new sentence should become more positive or more negative than the original sentence. 
For SSM, each suspected bug consists of a pair of questions with a derivation relation and the output of the tested object. 
Judges are required to determine if the two questions have the same semantics. 
If the suspected bug is related to the erroneous behaviors of the tested object, it is labeled $TP$; otherwise, it is labeled $FP$.
We present the suspected bugs in several tables and ensure that two judges manually review each suspected bug to enhance the reliability of the labeling results. 
For contested suspected bugs, we require all the judges to discuss them together to eliminate these discrepancies. 
For MRC, SA, and SSM, the Kappa coefficient of judgment results is 0.711, 0.656, and 0.778, respectively.
The labeling results of SA tend to have personal biases, so judges are also prone to disagreement. 
Thus, the judges are consistent in their labeling results for most of the suspected bugs.

\section{results and analysis}
This section introduces and analyzes the experimental results of \tool~in terms of testing efficiency and effectiveness, the authenticity of the generated tests, and the sources of FP and FN.
\subsection{The Efficiency of \tool}
According to the settings in Section~\ref{ref:settings}, we employ \tool~and CheckList to generate tests for three NLP tasks. 
Table \ref{tab:efficiency}~demonstrates the lengths of the seed sets and new testing sets generated by \tool~and CheckList, denoted as $\#Tests$, the total and average time cost of test generation of \tool~and CheckList, denoted as $TC_{total}$ and $TC_{avg}$, and the number of the reported bugs when inputting the tests generated by \tool~and CheckList into the two tested models, denoted as \#$R_{M1}$ and \#$R_{M2}$. 
For MRC, $M1$ and $M2$ are ALBERT and XLNet. 
For SA and SSM, $M1$ and $M2$ are ERNIE 2.0 and RoBERTa. 

\begin{table}[htbp]
  \caption{Test generation summary for each NLP task}
  \label{tab:efficiency}
  \centering
  \begin{tabular}{c|c|c|c|c}
  \toprule
   &&\textbf{MRC}&\textbf{SA}&\textbf{SSM}\\
   \hline
   \textbf{Original} &$\#Tests$&11,873&1821&40,430\\
   \hline
   \multirow{4}{*}{\textbf{\tool}}&$\#Tests$&850,426&96,609&262,462\\
   \cline{2-5}
    &$TC_{total}$&6043&8355&31,858\\
    \cline{2-5}
    &$TC_{avg}$&0.007&0.086&0.121\\
   \cline{2-5}
    &\#$R_{M1}$&46,858&3191&144,736\\
    \cline{2-5}
    &\#$R_{M2}$&22,942&2256&137,855\\
   \hline
   \multirow{4}{*}{\textbf{CheckList}}&$\#Tests$&683,752&93,835&246,296\\
   \cline{2-5}
    &$TC_{total}$&12837&18&493\\
   \cline{2-5}
   &$TC_{avg}$&0.020&0.0002&0.002\\
   \cline{2-5}
   \cline{2-5}
    &\#$R_{M1}$&28,597&211&41,607\\
    \cline{2-5}
    &\#$R_{M2}$&17,884&288&17,460\\
  \bottomrule
\end{tabular}
\end{table}


Based on Table~\ref{tab:efficiency}, we observe that using synonym substitution to mutate adjuncts is much faster than using the masked language model. 
This is because the main steps of synonym substitution include searching synonyms in the thesaurus and calculating the similarity of word vectors, which are less time-consuming than running the masked language model. 
In MRC testing, \tool~can generate an average of 71.6 tests when inputting one seed test due to the Cartesian product being applied, which is significantly more than the number of generated tests in SA and SSM testing. 
Compared with CheckList, \tool~reports much more suspected bugs when setting to generate a similar number of tests. 
For these three NLP tasks, the number of suspected bugs reported by \tool~is 1.5 times, 10.9 times, and 4.8 times that reported by CheckList, respectively. 
In the latter two tasks, the perturbation methods of CheckList consist of simple operations such as searching, concatenating, and replacing strings, which are less time-consuming than \tool~running the masked language model. 
Considering the significant number of suspected bugs reported by \tool, the average generation time between 0.007 and 0.121 is acceptable. 
Furthermore, due to the characteristics of generation methods, the types of bugs reported by CheckList also have limitations.
For software developers, CheckList reports similar types of bugs, which makes it meaningless to increase the number of reported bugs by generating more tests. 
We analyze this issue in more detail in Section~\ref{sec:effectiveness}. 
In contrast, the tests generated by \tool~cover more diverse sentence patterns and semantics. 
Thus, we can employ \tool~to evaluate NLP applications more comprehensively, as the possibility of detecting bugs and the types of bugs detected increases with the coverage of testing data. 

\subsection{The Authenticity of Generated Tests}
The authenticity of tests in our experiment refers to the grammatical and semantic correctness of the sentences contained in the tests. 
The more authentic a test is, the more likely it is to occur in real application scenarios.
We employ the sampling inspection to estimate the authenticity of tests generated by \tool. 
For each task, we randomly select 400 sentences from all the sentences generated by \tool~and divide them into four tables. 
Each judge is assigned to three tables corresponding to three tasks to complete the inspection. 
We employ the percentage of semantically grammatically correct sentences in all sentences as the evaluation metric of authenticity. 

For MRC, SA, and SSM, the percentage of semantically grammatically correct sentences generated by \tool~is 94.3\%, 95.3\%, and 97.0\%. 
The experimental results show that using the masked language model to mutate adjuncts can generate more authentic tests than synonym substitution. 
This is because the masked language model can better take contextual information into account, while the synonyms we searched occasionally have different usage or multiple meanings resulting in incorrect sentences.
The authenticity of tests in SA testing is least because the quality of colloquial movie reviews constituting the SST dataset is lower than the quality of sentences from other datasets.
Except for the quality of the seed, the accuracy of the derivation template and the suitability of mutated adjuncts also affect the authenticity of tests generated by~\tool. 
The components we select in the template generation and the adjunct mutation represent state-of-the-art levels in each field, and experimental results show that they are effective and reliable.
We do not check the tests generated by CheckList because their authenticity is strongly related to the seed sentences.
For example, injecting a pre-existing negation into a negative sentence or replacing a number that represents the special meaning in the seed sentence can lead to grammatical or semantic errors. 

\subsection{The Effectiveness of \tool}
\label{sec:effectiveness}
We employ the metric introduced in Section~\ref{sec:metric} to evaluate the effectiveness of \tool~in different application scenarios. 
Since we adopt manual evaluation, we define suspected bugs of MRC with the same problem and prediction as similar suspected bugs and screen the bugs detected by the seed set to reduce the labor and time cost of constructing the ground-truth. 
For SA and SSM, we adopt the sampling inspection, with each judge checking 100 suspected bugs. 
Table \ref{tab:effective}~demonstrates the number of detected bugs denoted as $\#Bugs$ and the precision of \tool~and CheckList denoted as $P_{bug}$. 
We highlight some examples of failed tests in Table \ref{tab:effective-mrc}. 
The green words represent the correct output, the red words represent the incorrect output of the tested object, and the blue words represent the differences between the test generated by \tool~and the seed. 

\begin{table}[htbp]
\caption{The number of detected bugs and the precision}
\label{tab:effective}
\centering
\begin{tabular}{c|c|c|c|c}
\toprule
\textbf{Task}& \textbf{Model} & \textbf{Metric}&\textbf{NLPLego}&\textbf{CheckList} \\
\hline
\multirow{4}{*}{\textbf{MRC}}&\multirow{2}{*}{ALBERT}&\#Bugs&1092&743\\
\cline{3-5}
&&$P_{bug}$&98.1\%&90.9\%\\
\cline{2-5}
&\multirow{2}{*}{XLNet}&\#Bugs&640&550\\
\cline{3-5}
&&$P_{bug}$&95.8\%&88.2\%\\
\hline
\multirow{4}{*}{\textbf{SA}}&\multirow{2}{*}{ERNIE 2.0}&\#Bugs&3086&144\\
\cline{3-5}
&&$P_{bug}$&96.7\%&68.2\%\\
\cline{2-5}
&\multirow{2}{*}{RoBERTa}&\#Bugs&2215&186\\
\cline{3-5}
&&$P_{bug}$&98.2\%&64.7\%\\
\hline
\multirow{4}{*}{\textbf{SSM}}&\multirow{2}{*}{ERNIE 2.0}&\#Bugs&135,328&36,905\\
\cline{3-5}
&&$P_{bug}$&93.5\%&88.7\%\\
\cline{2-5}
&\multirow{2}{*}{RoBERTa}&\#Bugs&126,551&16,727\\
\cline{3-5}
&&$P_{bug}$&91.8\%&95.8\%\\
\bottomrule
\end{tabular}
\end{table}

\begin{table*}[htbp]
\caption{Examples of failed tests for three NLP tasks}
\label{tab:effective-mrc}
 \centering
 \footnotesize
  \begin{tabular}{c|c|p{14.6cm}}
  \toprule
  \textbf{Task}&\textbf{Model}&\textbf{Example}\\
  \hline
  \multirow{4}{*}{\textbf{MRC}}&\multirow{2}{*}{\textbf{ALBERT}}&\begin{tabular}[l]{@{}l@{}}\textbf{Para:} ...Stated another \bluecode{manner}, \greencode{the instance is a particular input to the problem}, and \redcode{the solution} is the output corresponding to the given input. \\
  \textbf{Ques:} What is another name for any given measure of input associated with a problem?\\
  \textbf{Prediction:} \redcode{the solution} \textbf{Correct answer:} \greencode{instance}
  \end{tabular}\\
  \cline{3-3}
  &&\begin{tabular}[l]{@{}l@{}}\textbf{Para:} The uprising occurred a decade following \greencode{the death of Henry IV, a Huguenot before} \bluecode{changing} \greencode{to Catholicism, who had protected}\\\greencode{Protestants through the Edict of Nantes}. His successor \redcode{Louis XIII}, under the \bluecode{Regency} of his Italian Catholic mother Marie de' Medici...\\
  \textbf{Ques:} What King and former Huguenot looked out for the welfare of the group?\\
  \textbf{Prediction:} \redcode{Louis XIII} \textbf{Correct answer:} \greencode{Henry IV}
  \end{tabular}\\
  \cline{2-3}
  &\multirow{2}{*}{\textbf{XLNet}}&\begin{tabular}[l]{@{}l@{}}\textbf{Para:} Each year, the southern California area has about 10,000 earthquakes. \greencode{Nearly all of them are so small} that they are not \bluecode{sensed}. \redcode{Only}\\\redcode{several hundred are greater than magnitude 3.0}...\\
  \textbf{Ques:} Generally speaking, what size are the earthquakes that hit southern California?\\
  \textbf{Prediction:} \redcode{magnitude 3.0} \textbf{Correct answer:} \greencode{small}
  \end{tabular}\\
  \cline{3-3}
  &&\begin{tabular}[l]{@{}l@{}}\textbf{Para:} On 1 February 2007, the \bluecode{Eve} of the publication of IPCC's major report on \bluecode{clime}, a study was published suggesting that \greencode{temperatures and}\\\greencode{sea levels have been rising at or above the maximum rates} proposed during the last IPCC report in 2001. The study \redcode{compared IPCC 2001}\\  \redcode{projections on temperature and sea level change with} \bluecode{notices}... \\
  \textbf{Ques:} How did the 2001 IPCC report compare to reality for 2001-2006?\\
  \textbf{Prediction:} \redcode{compared IPCC 2001 projections on temperature and sea level change with notices}\\
  \textbf{Correct answer:} \greencode{temperatures and sea levels have been rising at or above the maximum rates}
  \end{tabular}\\
  \hline
    \multirow{4}{*}{\textbf{SA}}&\multirow{2}{*}{\textbf{ERNIE 2.0}}&\begin{tabular}[l]{@{}l@{}}
  A movie \redcode{pos} + tv \redcode{pos} $\rightarrow$ A tv movie \redcode{neg} \greencode{pos/neutral} \\
  \textbf{The increase of the} \redcode{neg}~\textbf{probability:} 0.828
  \end{tabular}\\
  \cline{3-3}
  &&\begin{tabular}[l]{@{}l@{}}
  One hour photo is an excellent snapshot of one man and his delusions; it's just too bad. \redcode{neg} + it doesn't have more flashes of insight \redcode{neg}\\$\rightarrow$ One hour photo is an excellent snapshot of one man and his delusions; it's just too bad it doesn't have more flashes of insight.\redcode{pos}\greencode{neg} 
  \\\textbf{The increase of the}\redcode{pos} \textbf{probability:} 0.870
  \end{tabular}\\
  \cline{2-3}
  &\multirow{2}{*}{\textbf{RoBERTa}}&\begin{tabular}[l]{@{}l@{}}
  It's a attempt.\redcode{neg} + brave\redcode{neg}\greencode{pos}$\rightarrow$ It's a brave attempt.\redcode{pos} \\
  \textbf{The increase of the}\redcode{pos} \textbf{probability:} 0.990
  \end{tabular}\\
  \cline{3-3}
  &&\begin{tabular}[l]{@{}l@{}}The latest installment in the pokemon franchise, pokemon 4ever is surprising less moldy and trite than the last two, however.\redcode{pos} + because\\much of the japanese series is set in a scenic forest\redcode{pos} $\rightarrow$ The latest installment in the pokemon franchise, pokemon 4ever is surprising less\\moldy and trite than the last two, however because much of the japanese series is set in a scenic forest.
  \redcode{neg}\greencode{pos} 
  \\\textbf{The increase of the}\redcode{neg} \textbf{probability:}  0.625
  \end{tabular}\\
  \hline
  \multirow{4}{*}{\textbf{SSM}}&\multirow{2}{*}{\textbf{ERNIE 2.0}}&\begin{tabular}[l]{@{}l@{}}
  Why did Trump purge members ? \redcode{=} \greencode{$\neq$}~Why did Trump purge members of his cabinet ? 
  \end{tabular}\\
  \cline{3-3}
  &&\begin{tabular}[l]{@{}l@{}}
  Why do folks ask questions on Quora?
  \redcode{=} \greencode{$\neq$}~Why do folks ask questions on Quora \textbf{that a Google search could answer faster and }\\\textbf{more quickly} ? 
  \end{tabular}\\
  \cline{2-3}
  &\multirow{2}{*}{\textbf{RoBERTa}}&
  \begin{tabular}[l]{@{}l@{}}
  Which country is most suitable for an Indian electronic engineer ? \redcode{=} \greencode{$\neq$}~Which country is most suitable for an Indian electronic engineer\\\textbf{to work and study} ?
  \end{tabular}\\
  \cline{3-3}
  &&
  \begin{tabular}[l]{@{}l@{}}
  How can I delete my Instagram account ? \redcode{=} \greencode{$\neq$}~How can I delete my Instagram account \textbf{if I don't know my own password or email} ? 
  \end{tabular}\\
  \bottomrule
\end{tabular}
\end{table*}


Since we employ the sampling inspection in SA and SSM testing, the number of detected bugs in SA and SSM testing is estimated based on the manual evaluation results. 
As observed from Table~\ref{tab:effective}, \tool~outperforms CheckList in all cases regarding the number of detected bugs and precision. 
While the difference in average precision is less pronounced in SSM testing than in other testing tasks, \tool~detects approximately 4.9 times as many bugs as CheckList. 
Due to the characteristics of generation methods of CheckList, it mostly mutates at the word level or splices a sentence unrelated to the context at the end of the seed sentence. 
In contrast to Checklist, one of the highlights of \tool~is to endeavor to exploit the potential of the seed test by inserting words/phrases/clauses before, during, or at the end of seed sentences in a reverse-disassembly manner. 
Therefore, the quality of tests generated by \tool~is higher than CheckList, and the types of bugs detected are less homogeneous. 
We find that 90\% of failed SA tests generated by CheckList are related to injecting negation. 
However, injecting negation into a sentence stating an objective fact does not mean that the sentiment of the sentence can change significantly, resulting in its precision being only 66.5\%. 
Meanwhile, in SSM testing, most of the bugs detected by CheckList are caused by the incapability of the NLP application to distinguish names, place names, or numbers, which are all very similar. 
For software developers, these bugs are less valuable than the bugs detected by \tool, which are caused by changes in sentence patterns and semantics. 

\subsection{False Positives and False Negatives}
The type of false positives is annotated by judges while checking the suspected bugs.
False negatives usually occur when the Oracle information of the test is unclear. 
We employ the correct answers given by SQuAD 2.0 when generating the tests of MRC and obtain the correct labels based on the metamorphic relation when generating the tests of SSM.
So there are no false negatives when testing MRC and SSM. 
We explore the source of false negatives through random inspection in SA testing because we do not have a way to check all the generated tests. 

We observe that the three tasks have a common source of false positives: the presence of grammatical or semantic errors in the tests causing false positives.
For example, if the words in fixed collocations are replaced by synonyms, such as ``to date'' becomes ``to appointment'', the semantics of the sentence will significantly change or even become very strange. 
Or \tool~occasionally loses ``the'' word before the plural noun when generating the template, causing "the" to be perceived as an adjunct. 
The semantics of the sentence does not change before and after inserting ``the'',  resulting in false positives in SSM testing. 
For MRC, the other sources of false positives include: the semantics of the sentence has changed, causing the answer to change (55\% of the total), and the answer is correct but not present in the ground-truth set (9\% of the total). 
For SA, another main source of false positives is the presence of negation in a minority of sentences, leading to the reverse sentiment change after inserting an adjunct. 
For SSM, another main source of false positives is that the semantics of the question does not change significantly after inserting an adjunct, just like the question ``Is it true that pilots do jobs after investing in their CPL?'' and the question ``Is it true that pilots do jobs after investing in their CPL (commercial pilot program)?''.
By inspecting the outputs randomly in SA testing, we notice that \tool~can not detect instances where both the sentiment analysis result of the newly generated sentences and the inserted adjunct are erroneous. 

\subsection{Discussion}
The assembling test generation method implemented in \tool~and the test generation method applied in \textit{Skeletal Program Enumeration (SPE)}~\cite{zhang2017skeletal} both employ the idea of template-based generation. 
The test generation method applied in SPE consists of two main steps: skeleton generation and program enumeration components. 
The skeleton generation and filling methods are different from our assembling test generation method in the following aspects: 1) The types of input and output data are different. 
Less structured natural language is more difficult to process than highly structured code. 
2) \tool~assembles grammatical elements step by step in a specific order rather than filling templates all at once. 
The code skeleton can not run without being fully filled, while the natural language is more flexible and can be generated gradually to create the conditions for conducting metamorphic testing. 

The key to \tool~being a general-purpose testing tool is that the intergenerational relationships formed by the assembling step can be employed flexibly to construct various MRs. 
Through experiments, we confirm that \tool~is an effective general-purpose testing tool for NLP applications. 
The core idea of \tool~is to abstract the concrete test and then gradually assemble the components of the concrete test based on the abstraction result. 
While CheckList supports users to manually create templates, one of the highlights of \tool~is to automate the converting process between abstraction and concretization with high accuracy. 
Even without the adjunct mutation strategies, the conversion from concretization to abstraction can be considered as a word/phrase/clause level sentence mutation method. 
\tool~breaks through the limitations of existing methods that mostly focus on replacing specific words, splicing context-independent text elements, or introducing spelling errors.
Thus, \tool~can make the most of limited resources to expand the coverage of tests.
We believe that this idea of interchangeability between concretization and abstraction can be applied to the automated testing of other types of software. 

We notice from Table \ref{tab:effective-mrc}~that the language comprehension and word discrimination capabilities of NLP applications need to be further strengthened. 
Even if the sentence semantics are not changed, only changing individual words can cause the MRC applications to locate the key sentence incorrectly. 
The MRC applications are more prone to errors when the keywords appear in multiple sentences, which can reveal that the MRC applications use the memorizing keyword trick. 
For SA and SSM testing, we observe that the SA and SSM applications sometimes can not distinguish and understand the meanings of words in different contexts and the logical relations in sentences.
Due to the distribution bias of the training data, SA applications often identify neutral nouns as words containing sentiment and incorrectly assign stereotypes to adjectives or adverbs without considering the context. 
In SSM testing, the SSM applications can not discern sentences that are structurally similar but semantically different, indicating that the SSM applications may not really understand the sentence semantics but rather determine whether the sentences are semantically similar by the degree of similarity of the included words. 
\subsection{Threats to Validity}
\subsubsection{Threats to internal validity.} 
The threats to internal validity of \tool~mainly come from the impact of support components, the testing module selection, and the ground-truth construction. 
The support components of \tool~are the sentence compression model \textit{SLAHAN}, the masked language model \textit{BERT}, and the NLP tools \textit{Stanford CoreNLP} and \textit{spaCy}. 
Although advanced in their respective fields, they are still not 100\% accurate. 
It is inevitable that the tests generated by \tool~can not be 100\% semantically and syntactically correct. 
After experiments, we confirm that the correctness of generated tests is between 94.3\% and 97\%, which is acceptable for an automated test generation tool. 
A complete NLP application consists of multiple modules (the input module, the pre-processing module, the NLP model, etc.). 
We select the NLP model as the tested object because it is the module most closely associated with the performance of NLP applications. 
Also, we train the judges and provide explanatory documents before the ground-truth construction to reduce the negative impact of manual evaluation on the experiment results. 
\subsubsection{Threats to external validity.} 
We mainly consider the threats to external validity are whether the performance of \tool~can maintain after applying it to other types of NLP applications. 
The key to applying \tool~is constructing the MR, and the characteristics of the sentence generated by \tool~make constructing the proper MR easier. 
For example, when testing machine translation software, sentences with the derivation relation can be employed to check structure or semantic invariance between their translation results. 
When testing intelligent dialogue systems, we can detect bugs by checking for the intent or answer invariance. 
Thus, we think that \tool~can be applied well to testing other types of NLP applications. 

\section{Related Work}
\textbf{Text Data Augmentation.} To address issues such as data scarcity and lack of data diversity, researchers have proposed data augmentation strategies, which are usually used in the training and testing of NLP models~\cite{li2022data,shorten2021text}. 
Li et al. divided commonly used text data augmentation methods into three categories: paraphrasing-based methods, noising-based methods, and sampling-based methods~\cite{li2022data}. 
The paraphrasing-based methods aim to generate data that contain similar information to the original data, such as back-translation~\cite{zhang2020parallel}, synonym substitution~\cite{zhang2015character}, and semantic embeddings~\cite{regina2020text}. 
The noising-based methods inject faint noise that does not seriously affect the semantics of the original data, such as word-level random swap or deletion~\cite{wei2019eda} and sentence insertion~\cite{yan2019data}. 
The sampling-based methods are similar to the paraphrasing-based methods, with the difference that they are task-specific (require task information like labels and data format) and depend on artificial heuristics~\cite{min2020syntactic,kober2020data}. 
According to the above definition, our method can be classified into the third category. 
Unlike the task-specific methods, experimental results have confirmed that this method can be flexibly applied to different NLP tasks. 

\noindent \textbf{Testing Methods for NLP Applications.} Currently, most of the testing methods for NLP applications are designed and implemented based on the characteristics of NLP tasks. 
For example, He et al. proposed a structure-invariant testing method for machine translation software in 2020, which detects translation errors by checking whether the outputs of the MT software satisfy structure invariance~\cite{he2020structure}. 
In 2021, Chen et al. proposed a metamorphic testing method for MRC software, including seven metamorphic relations specifically designed for MRC (e.g., invertion with the antonymous adjective, invertion with tense change and invertion with order change, etc.)~\cite{chen2021validation}. 
The task-specific testing methods mentioned above all design a proper metamorphic relation first and then determine the corresponding test generation method. 
Thus, the test generation methods they employed are not generalizable to other NLP tasks.
Marco et al. proposed a behavior testing tool, CheckList, for various NLP applications, which is more similar to \tool.
It contains some widely-used test generation methods to validate whether the NLP applications follow some generally accepted basic properties. 
Although CheckList mentions the test generation with templates, unlike \tool, which can automatically create templates and tests, it requires users to manually confirm templates to ensure the availability of the tests. 

\section{conclusion}
In this paper, we propose the assembling test generation method for NLP applications and implement it into \tool. 
\tool~disassembles the seed sentence into the template and adjuncts and then generates new sentences by assembling context-appropriate adjuncts with the template in a specific order. 
Unlike the task-specific tool, the tests generated by \tool~have derivation relations and different degrees of variation, which help meet the testing requirements of various NLP applications. 
To validate \tool, we experiment with three common NLP tasks, i.e., MRC, SA, and SSM, identifying failures in four widely-used advanced models. 
Given seed tests from SQuAD 2.0, SST, and QQP, \tool~successfully detects 1,732, 5301, and 261,879 incorrect behaviors with around 95.7\% precision in three tasks, respectively.
The experiment results show that \tool~can efficiently generate tests for multiple NLP tasks to detect erroneous behaviors of NLP applications effectively. 

\bibliographystyle{ACM-Reference-Format}
\bibliography{ref.bib}
\end{document}